# An AI-powered blood test to detect cancer using nanoDSF

Philipp O. Tsvetkov[1,2], Rémi Eyraud[3,4], Stéphane Ayache[3], Anton A. Bougaev[5], Soazig Malesinski[1], Hamed Benazha[3], Svetlana Gorokhova[6,7], Christophe Buffat[8,9], Caroline Dehais[10,11], Marc Sanson[10,11], Franck Bielle[10,12], Dominique Figarella-Branger[1,13], Olivier Chinot[1,14], Emeline Tabouret[1,14], François Devred[1,2]

[1] Aix-Marseille Univ, CNRS, UMR 7051, INP, Inst Neurophysiopathol, Faculté des Sciences Médicales et Paramédicales, Marseille, France

[2] Aix-Marseille Univ, PINT, Plateforme Interactome Timone, Faculté des Sciences Médicales et Paramédicales, Marseille, France

[3] Aix Marseille Université, CNRS, LIS, Marseille, France

[4] Univ Lyon, UJM-Saint-Etienne, CNRS, Laboratoire Hubert Curien UMR 5516, France

[5] Oracle Labs, San Diego, CA, 92121, USA

[6] Faculté des Sciences Médicales et Paramédicales, Marseille Medical Genetics, Aix Marseille Université, INSERM, Marseille, France.

[7] Service de génétique Médicale, Hôpital de la Timone, APHM, Marseille, France.

[8] APHM, Biochemistry and Endocrinology, Hôpital de la Conception, Marseille, France.

[9] Aix-Marseille Université, MEPHI, IRD, APHM, Marseille, France.

[10] Sorbonne Université, Inserm, CNRS, UMR S 1127, Institut du Cerveau et de la Moelle épinière, ICM, Paris, France.

[11] AP-HP, Hôpitaux Universitaires La Pitié Salpêtrière - Charles Foix, Service de Neurologie 2-Mazarin, Paris, France

[12] AP-HP, Hôpitaux Universitaires La Pitié Salpêtrière - Charles Foix, Département de Neuropathologie, Paris, France

[13] APHM, CHU Timone, Service d'Anatomie Pathologique et de Neuropathologie, Marseille, France

[14] Service de Neuro Oncologie, Hopital de La Timone, APHM, Marseille, France

## *Abstract*

We describe a novel cancer diagnostic method based on plasma denaturation profiles obtained by a non-conventional use of Differential Scanning Fluorimetry. We show that 84 glioma patients and 63 healthy controls can be automatically classified using denaturation profiles with the help of machine learning algorithms with 92% accuracy. Proposed high throughput workflow can be applied to any type of cancer and could become a powerful pan-cancer diagnostic and monitoring tool from a simple blood test.

## *Article*

Diffuse gliomas are the most frequent and aggressive primary brain tumors in adults. Currently, no curative treatment is available despite the association of surgical resection, radiotherapy and chemotherapy as first-line treatment [1]. Another major challenge in glioma patient management is obtaining timely and precise histological and molecular characterization of the tumor in order to establish diagnosis and orient treatment. However, biopsies of these tumors are often impossible due to their deep or diffuse location, or due to patient comorbidities. In these cases, treatment is chosen based on MRI neuro-imaging characteristics that are often insufficient. More efficient ways to follow patients are also needed. Indeed, the evaluation of patients under treatment is currently based on MRI, steroid dose and clinical examination, which are often difficult to interpret after radiotherapy, anti-angiogenic therapy or immunotherapy [2,3]. Finally, an accurate and timely detection of the disease recurrence is crucial to optimize the therapeutic options and to improve patients' treatment and quality of life. Thus, there is an urgent need in the neuro-oncology field to design new easy-to-use methods that are less invasive than histological examination and more efficient than neuroimaging in order to help patient diagnosis and to follow disease progression.

We have previously shown that despite the blood-brain barrier, the presence of glioblastoma induces changes in patients' plasma that can be detected by Differential Scanning Calorimetry (DSC) [4,5]. We now describe a novel method for high-throughput plasma profiling by repurposing another fundamental research method: nanoDSF (Differential Scanning Fluorimetry). NanoDSF, which was originally designed to study protein thermostability [6,7], is based on the modifications of the intrinsic fluorescence of the macromolecules upon their thermal denaturation. In this study we applied nanoDSF to analyze the plasma of patients affected by glioma and compared the obtained denaturation profiles to that of healthy individuals.

We conducted this study on a bicentric cohort of 84 glioma patients with a median age of 49.3 years (range, 19.6 - 77.5). Twenty-two patients (26%) presented with a 1p/19q codeleted IDH mutated oligodendroglioma, 25 patients (31%) with an IDH mutated astrocytoma and 37 patients (43%) with an IDH wild-type astrocytoma (see Table 1 for more detailed patient characteristics). Plasma samples from this cohort and from the 63 healthy controls were loaded to 24-capillary chips and then scanned using nanoDSF Prometheus NT.Plex instrument (Nanotemper) in order to obtain the denaturation profiles in the range from 15 to 95°C. Raw data were exported into datasets that contained all of the nanoDSF outputs: fluorescence at 330 and 350 nm (F330 and F350), the ratio of these values (F330/F350) as well as absorbance at 350 nm (A350). The first derivatives of F330/F350 were plotted to visualize denaturation (Figure 1). As seen from this figure, the mean denaturation profile of the glioma patients' plasma was drastically different from that of healthy individuals.

The observed difference in the plasma denaturation profiles between the glioma and the healthy samples can be explained by the variation in thermal stability of the plasma constituents. Indeed, the plasma denaturation profiles correspond to the cumulative sum of those from the most abundant plasma proteins [8]. Since the thermal denaturation profile of a protein is an intrinsic property dependent on its structure, modifications such as mutation, post translational modifications, or ligand binding can significantly impact this profile. Similarly to results obtained with DSC [5,8], we observed low variability in nanoDSF denaturation profiles within controls, regardless of sex or age of the individuals. This can be explained by the fact that the composition of many biofluids such as plasma, serum, cerebrospinal fluid, is meticulously maintained by the organism, thus resulting in a reproducible denaturation profile. Such healthy plasma equilibrium is altered in glioma patients, leading to the emergence of a different glioma profile. Even though further studies are needed to identify the molecular basis of plasma changes in glioma patients, observed differences can be used for diagnostic purposes.

In order to differentiate between the denaturation profiles of healthy individuals and glioma patients, we set out to design an automated way to classify the obtained profiles using an Artificial Intelligence (AI) approach. Moreover, automation is needed for future applications of this approach to much larger scale analysis in clinics as well as to detect possible subtle differences between subgroups of samples. We tested several machine learning (ML) algorithms [9] on the raw data: the classical Logistic Regression (LR), the often well-performing Support Vector Machine (SVM), the well-known Neural Networks (NN), and two different ensemble methods: Random Forest (RF) and Adaptive Boosting (AdaBoost). These algorithms were evaluated using a leave-one-out approach where each datum is used once as a test, while the others are used to train the automatic classifiers: the obtained values were thus averages on as many experiments as there are data. The three nanoDSF outputs (F330, F350 and A350) were tested independently and in combination as input for these artificial intelligence algorithms. Table 2 shows the results obtained with the five ML algorithms (LR, SVM, NN, RG and AdaBoost) using the settings allowing the best observed performance on our 147 samples (84 patients and 63 controls). NN and AdaBoost algorithms had the best accuracy of above 92%, while all others achieved around 90% of correct classification. LR algorithm provided the lowest number of false positives (four healthy individuals were wrongly classified as glioma patients), while AdaBoost was better at reducing the number of false negatives (five glioma patients classified as healthy). The two algorithms with highest accuracy (NN and AdaBoost) had closely related small numbers of the two error types (false positives: 5 and 6 respectively, false negatives: 6 and 5 respectively). When the algorithm allowed it, we also tested its version that focuses on minimizing the number of false negatives, in order to decrease the possibility of missing the diagnosis of glioma that could have devastating consequences given the rapidly developing nature of this disease. As seen from the Table 2, the false negative focussing version of the SVM algorithm (fnf-SVM) maintained the same level of overall accuracy as the original SVM (87.07% of correct classification) while obtaining just one false negative (corresponding to 1.19% of glioma patients mis-classified by this algorithm). Taken together, our results show that detection of glioma based on the denaturation profile of plasma can be efficiently automated. Moreover, combining high-throughput nanoDSF and automated data treatment by machine learning makes our approach compatible with large-scale applications in clinics for cancer detection using a simple blood test.

Detection of cancers by a minimally invasive blood test, or "liquid biopsy", has been a long-thought goal in the field. A number of different cancer detection methods have been tested over the past ten years, which are based on Differential Scanning Calorimetry (DSC) [8], infrared technology (ATR-FTIR) [10] as well as on the detection and isolation of cell-free nucleic acids, extracellular vesicles and circulating tumor cells [11–14]. Among these, many pilot studies have previously tested DSC of biofluids as a one-step and low cost approach for diagnosis of a great number of diseases including several types of cancers [8,15,16], raising hopes about creation of a unique pan-cancer diagnostic tool. However, despite the efforts invested in developing this approach, technical restrictions and low throughput of DSC instruments made them impossible to be transferred for wide use in clinics. In our study, we describe a major technical breakthrough allowing us to overcome these obstacles. Indeed, our approach using the nanoDSF instrument requires minimal quantity of plasma, no need for sample preparation, and allows much faster sample handling due to disposable capillaries and high-powered fully automated data analysis using machine learning algorithms. Compared to classical DSC, our method provides a significant increase in throughput and reproducibility while decreasing the possibility of technical error.

In conclusion, our proof-of-concept study demonstrates the possibility to automatically distinguish glioma patients from healthy controls by a simple blood test, using a novel technology that combines differential scanning fluorimetry and machine learning algorithms. We propose that plasma profiling using denaturation signatures by nanoDSF can be used to develop low cost and high throughput diagnostic methods for cancers and human disease in general.

| FACTORS | N | % |
|---|---|---|
| **Age** (median, min-max) | 49.3 ( 19.6 - 77.5) | |
| **Gender** (Men/women) | 45/39 | 54/46 |
| **KPS** (median, min-max) | 80 (50 - 100) | |
| 50 | 1 | 1 |
| 60 | 10 | 13 |
| 70 | 17 | 21 |
| 80 | 23 | 29 |
| 90 | 20 | 25 |
| 100 | 9 | 11 |
| Steroids | 51 | 64 |
| **Histology** | | |
| *Oligodendroglioma* | | |
| Grade II | 2 | 2 |
| Grade III | 20 | 24 |
| *Astrocytoma IDHmut* | | |
| Grade II | 3 | 4 |
| Grade III | 19 | 23 |
| Grade IV | 3 | 4 |
| *Astrocytoma IDHwt* | | |
| Grade II | 7 | 8 |
| Grade III | 11 | 13 |
| Grade IV | 19 | 22 |
| **Surgery** | | |
| Gross total resection | 33 | 41 |
| Partial resection | 48 | 59 |
| **Adjuvant treatment** | | |
| Radiotherapy alone | 10 | 12 |
| Chemotherapy alone | 12 | 14 |
| Radiotherapy + chemotherapy | 57 | 68 |
| None | 5 | 6 |

**Table 1. Glioma patient characteristics**

|  | Original algorithms | | | | | FN minimized |
| --- | --- | --- | --- | --- | --- | --- |
| Algorithm | LR | SVM | NN | RF | AdaBoost | fnf-SVM |
| Accuracy | 89.80 | 87.07 | **92.52** | 89.12 | **92.52** | 87.07 |
| False positives | **4** | 13 | 5 | 8 | 6 | 18 |
| False negatives | 11 | 6 | 6 | 8 | **5** | 1 |

**Table 2. Best obtained results with the different machine learning algorithms.**

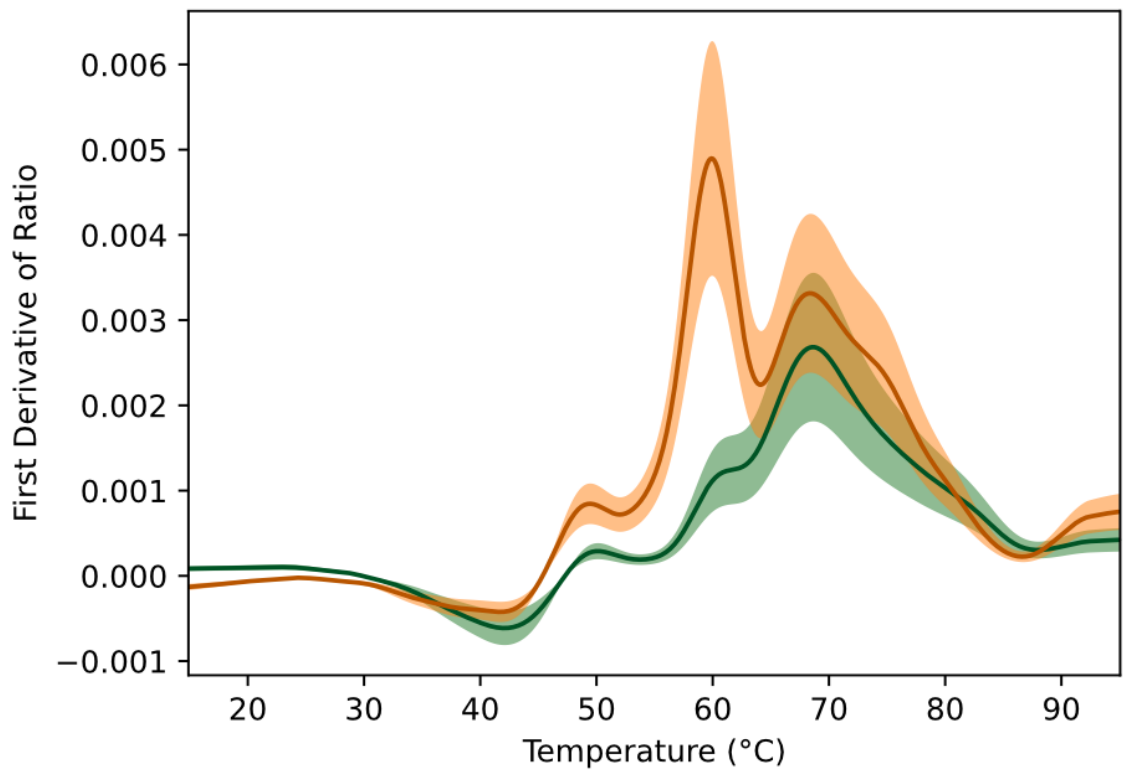

**Figure 1. Means of the first derivatives of the ratio F330/F350.** The red curve corresponds to glioma patients while the green one is the one of the controls. The color regions graphically shows the variability of the data by indicating the confidence interval at 99%.

*Methods section*

**Patients**

Patient cohort consisted of 51 patients included at La Timone Hospital (Marseille) from June 2009 to February 2017 and 33 patients included at La Pitié Salpétrière Hospital (Paris) from November 2008 to September 2016. Eligible patients included were those aged 18 years or older with newly diagnosed glioma for whom plasma samples were available at the time of diagnosis, before adjuvant treatment. Clinical evaluations were performed every cycle and imaging evaluations were performed every two cycles. Treatment responses and disease progression were reviewed using the RANO criteria [3]. All patients provided written informed consent in accordance with institutional, national guidelines and the Declaration of Helsinki.

**Plasma samples**

Blood samples from this cohort and from 63 healthy controls were collected into EDTA tubes, separated by centrifugation (2000g, 10 minutes, 20°C, twice) within 30 minutes and then stored at - 80°C. No other specific purification step was added in order not to perturb the interactome or alter the chemical state of plasma proteins. Before nanoDSF analysis, samples were thawed rapidly at 37°C, centrifuged and loaded on a 10 µL capillary.

**Sample analysis by nanoDSF**

Plasma samples were loaded to 10 µL capillaries and scanned using nanoDSF Prometheus NT.Plex instrument (Nanotemper) at 5% of laser power and 1°C/min heating rate to obtain denaturation profiles in the range from 15 to 95°C. The machine can analyse 24 samples at once: we carefully mixed patients and controls to avoid any batch effect. Raw data were exported into datasets that contained all the nanoDSF outputs : fluorescence at 330 and 350 nm (F330 and F350) as well as the ratio of these values (F330/F350) and absorbance at 350 nm (A350).

*Algorithm trainings*

The code used was written in Python. The data preparation was done using the pandas library (https://pandas.pydata.org) while the machine learning algorithms were run using the scikit-learn toolbox (https://scikit-learn.org). Raw data from the nanoDSF instrument (F330, F350 and A350) were interpolated using InterpolatedUnivariateSpline from the scipy.interpolate module in order to ensure the same temperature alignment for all data. The different tested implementations are: (1) LogisticRegression from the linear_model module with parameter max_iter sets to 1000; (2) SVC from the svm module with the following combination of parameters: kernel='poly', gamma='auto', C=1, degree between 1 and 3; kernel='rbf', C=1, gamma within [0.001, 0.01, 0.1, 1, 10]. The fnf-SVM results were obtained using SVC with the same parameters except for class_weight that was set to {0:1, 1:100} (instead of the default None value); (3) MLPClassifier from the neural_network module with different architecture (reported results correspond to 3 hidden layers of 750, 200, 50, respectively), max_iter was fixed to 5000 and learning_rate='adaptive'; (4) RandomForestClassifier from the ensemble module with parameter n_estimators fixed to 500; (5) AdaBoostClassifier from the module ensemble with a DecisionTreeClassifier from the module tree as weak classifier (parameter base_estimator) with max_depth taken between 1 and 3, n_estimators set to 100. All algorithms were evaluated using the split from the LeaveOneOut method of the model_selection module.

## *Acknowledgments :*

This study was supported by research funding from the Cancéropôle Provence-Alpes-Côte d'Azur, the French National Cancer Institute (INCa), Provence-Alpes-Côte d'Azur Région, MIC grant from ITMO Cancer of Aviesan, Patient association ARTC Sud and by INCa-DGOS-Inserm_12560 grant (SiRIC CURAMUS). We would also like to thank AP-HM Tumor Bank (authorization number 2017-2986) and the Onconeurotek Tumor Bank (APHP).